\documentclass{raa}
\usepackage{graphicx,times}
\usepackage{natbib}
\usepackage{amssymb,amsmath}
\bibpunct{(}{)}{;}{a}{}{,}

\begin{document}

   \title{Periodic variation and phase analysis of grouped solar flare with sunspot activity$^*$
\footnotetext{\small $*$ Supported by the National Natural Science Foundation of China.}
}

 \volnopage{ {\bf 20XX} Vol.\ {\bf X} No. {\bf XX}, 000--000}
   \setcounter{page}{1}

   \author{ Hui Deng\inst{1,2}, Ying Mei\inst{1}
   ,Feng Wang\inst{*1,2,3},}

   \institute{Center For Astrophysics, Guangzhou University, Guangzhou~510006, China; {\it fengwang@gzhu.edu.cn}\\
        \and
             CAS Key Laboratory of Solar Activity, National Astronomical Observatories, Beijing~100012, China\\
        \and
             Kunming University of Science and Technology, Kunming~650051, China\\
\vs \no
   {\small Received 20XX Month Day; accepted 20XX Month Day}
}

\abstract{Studies on the periodic variation and the phase relationship between different solar activity indicators are useful for understanding the long-term evolution of solar activity cycle. Here we report the statistical analysis of grouped solar flare (GSF) and sunspot number (SN) during the time interval from January 1965 to March 2009. We find that, 1) the significant periodicities of both GSF and SN are related to the differential rotation periodicity, the quasi-biennial oscillation (QBO), and the eleven-year Schwabe cycle (ESC), but the specific values are not absolutely identical; 2)  the ESC signal of GSF lags behind that of SN with an average of 7.8 months during the considered time interval, implying that the systematic phase delays between GSF and SN originate from the inter-solar-cycle signal. Our results may provide evidence about the storage of magnetic energy in the corona.
}

   \authorrunning{F. Wang et al. }            %author_head in even pages
   \titlerunning{Periodic variation and phase analysis of solar activity cycle}  % title_head in odd pages
   \maketitle

%________________________________________________ sections below
%
\section{Introduction}           %% first-level sections will be auto-capitalized
\label{sect:intro}

Understanding the dynamic processes of solar interior and atmosphere and how they produce the long-term cyclic variation continues to challenge the solar communities \citep{2017SSRv..210..367C}. The available solar databases provide a broader chance, showing a wide variety of periodic behavior, phase asynchrony, hemispheric coupling, chaotic behavior, fractal property as well as cycle amplitude and length \citep{2016AJ....151....2D,2016AJ....151...70D}. Furthermore, advances in data analysis techniques, such as time-frequency analysis and machine deep learning, are inspiring studies which allow us to further explore the nonlinear dynamics of the Sun in far greater detail \citep{2012AdSpR..50.1425D}. 

During the past few decades, the statistical relationship between solar flare activity and sunspot number (or sunspot area) is an interesting topic. \cite{1988AdSpR...8...67W} reported that solar flare occurrence and background flux in the soft X-ray (SXR) wavelength are delayed with respect to sunspot activity, and the relative phase shifts are around two to three years between the peak times during solar cycle 21. Because the SXR flare occurrence and background flux are considered to be dominated by the post-flares emission from the dominant active regions, so \cite{1994SoPh..152...53A} interpreted that the phase delay is possibly due to the increasing complexity of coronal magnetic structures in the decay phase of the solar cycle. By studying the temporal variation of SXR background flux and its relation to the flaring rate, energetic event rate, and the solar cycle, \cite{1993JGR....9811477W} did not find any evidence for the phase delay between SXR flare occurrence and sunspot activity during solar cycle 22 (from January 1986 to May 1992). If the solar cycles 21 and 22 are considered together, the average phase delay derived from the correlation analysis is found to be 6 months, as studied by \cite{2001ApJ...557..332W}.

To describe how the magnetic free energy in the solar corona varies in response to variations in the supply of energy to the system and to changes in the flaring rate, \cite{2001ApJ...557..332W} presented a detailed model for dynamic energy balance in the corona over the solar cycle. Their model predicted that both the flaring rate and the free energy of the system lag behind the driving of the system with a lag of around 11 months. To test the results of their model, \cite{2003SoPh..215..111T} analyzed the temporal evolution of solar flare occurrence with respect to sunspot activity during solar cycles 19-23. They found that, for solar cycles 19, 21, and 23 (namely, odd solar cycles), a characteristic phase lag between flare activity and sunspot activity is in the range from 10 to 15 months, which is consistent with the model predictions by \cite{2001ApJ...557..332W}. However, no characteristic phase lag larger than zero is found for solar cycles 20 and 22 (namely, even solar cycles). \cite{2013BASI...41..237F} studied the phase relationship between grouped solar flare and sunspot number by several time-frequency analysis methods, they found that the phase relationship between the two is not only time-dependent but also frequency-dependent, implying that their relationship is a complex nonlinear relationship. The relationships between solar flare parameters (such as the total importance, the time duration, the flare index, and the flux) and sunspot activity as well as those between geomagnetic activity (aa index) and the flare parameters was well studied by the \cite{2012RAA....12..400D}, they found that their relationship can be well described by an integral response model with the response time scales of about 8 and 13 months, respectively. That is, solar flare is considered to be related to the accumulation of solar magnetic energy in the past through a time decay factor, and  it will help the researchers to understand the mechanism of solar flares and to improve the prediction of the solar flares. \cite{2011AnGeo..29.1005D} proposed an integral response model to describe the relationship between geomagnetic activity and solar activity (represented by sunspot number, and the proposed model can naturally explain some phenomena related to geomagnetic activity and solar activity, such as the phase lag between flare activity and sunspot number.

The periodic scales of solar activity indices have a broad ranges, varying from several days to tens of years. Except for the 27-day rotation period and the 11-year Schwabe cycle, many other quasi-periodicities are found in the past seven decades. For example, \cite{2010SoPh..264..255K} studied the periodicities of solar flare index during solar cycles 21-23, and found that a lesser number of periodicities is found in the range of low frequencies (long periods) while the higher frequencies display a great number of periodicities. They also found that the periodicities of solar activity in different solar cycles are not identical. For the temporal features between solar flare activity and sunspot number or area, the clear relationships are not fully studied.  For instance,  we do not know whether the periodicities of flare occurrence and sunspot activity are identical, and whether the phase relationship between flare activity and sunspot activity depends on the considered periodicities. Therefore, it is needed to study the quasi-periodic variations and the phase relationship between solar flare occurrence and sunspot activity. 

In this paper we report the periodic variation and phase relationship between grouped solar flare and sunspot numbers during the time interval from January 1965 to March 2009 (soalr cycles 20-23). Two nonlinear time-frequency analysis techniques, namely the ensemble empirical mode decomposition (EEMD) and the cross-recurrence plot (CRP), are applied in this work. In Section 2, the data sets and the analysis methods are introduced. The statistical analysis results are shown in Section 3. And finally, the main conclusions are summarized in Section 4.

\section{Data sets and analysis methods}
\label{sect:Obs}

\subsection{Observational Time Series}

The GSF time series is publicly downloaded from  the website of the National Geophysical Data Center (NGDC)\footnote{ftp://ftp.ngdc.noaa.gov/STP/SGD/sgdpdf/Number\_of\_Solar\_Flares.pdf}. The time interval of GSF data set is from January 1965 to March 2009, almost covering solar cycles 20-23. The term ``grouped" means the observations of the same flare event by different solar sites were lumped together and counted as one. This indicator is thus different from the classical flare activity indices, such as H$\alpha$ flare index, soft X-ray flare, and hard X-ray flare \citep{2007JGRA..112.5105G}.

The SN data set is freely obtained from the World Data Center (WDC) --- Sunspot Index and Long-term Solar Observations (SILSO), Royal Observatory of Belgium, Brussels\footnote{http://www.sidc.be/silso/datafiles}. The SN time series (version 2.0) begins from January 1749 to April 2019 and is updated every month \citep{2014SSRv..186...35C}. Here, the time period from January 1965 to March 2009, the common time interval to the GSF data, is extracted.

Figure 1 displays the monthly counts of GSF (upper panel) and SN (lower panel) during the time period from January 1965 to March 2009, covering the solar cycles 20-23. As the figure shows, the temporal evolution of GSF obviously differs from that of SN, indicating that they exhibit different features.

\begin{figure}
   \centering
   \includegraphics[width=14.0cm, angle=0]{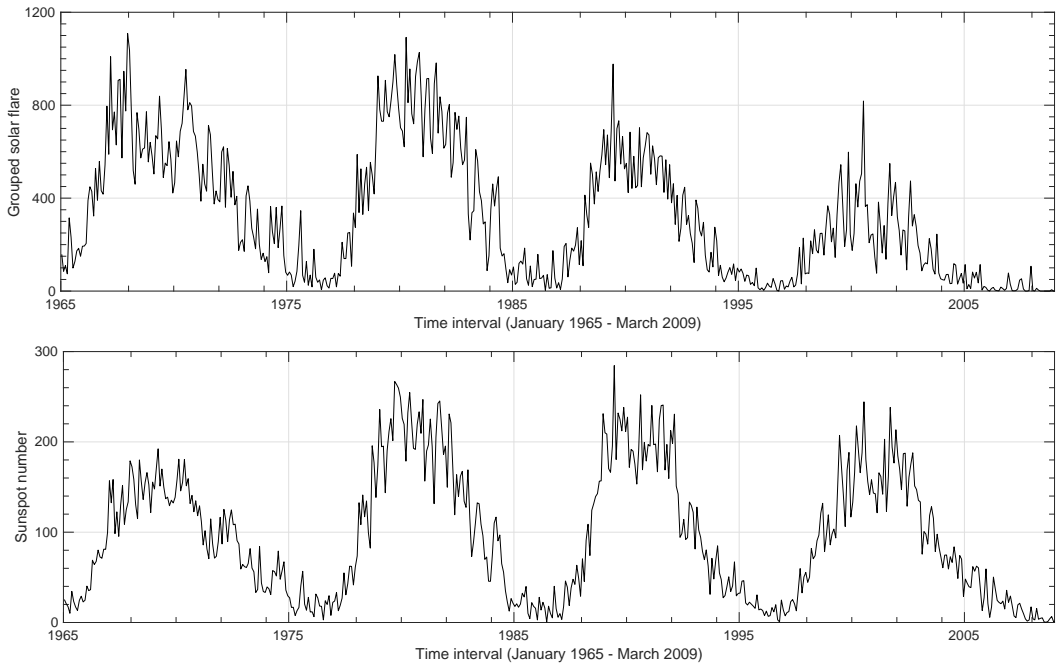}
  % \begin{minipage}[]{85mm}
   \caption{Monthly counts of grouped solar flare (upper panel) and sunspot number (lower panel) during the time interval from January 1965 to March 2009.}
%\end{minipage}
   \label{Fig1}
   \end{figure}

\subsection{Methods of Analysis}

\subsubsection{Ensemble Empirical Mode Decomposition}

Contrary to the traditional data decomposing techniques, the empirical mode decomposition (EMD) is an empirical, intuitive, and adaptive method, without requiring any predetermined basis functions. This time-frequency analysis method was first proposed by \cite{1998RSPSA.454..903E} and has been applied in many research fields. For solar physics studies, \cite{2012ApJ...747..135L} applied it to decompose the solar constant into three components: the rotation signal, the annual-variation, and the inter-solar-cycle signal; to study the periodic components of polar faculae and coronal green line intensity, \cite{2014AdSpR..54..125D} and \cite{2015JASTP.122...18D} used the EMD to reveal the long-term temporal variation of solar time series; \cite{2016AJ....151...76X} used it to extract the intrinsic mode functions (IMFs) of the solar mean magnetic field observed at the Wilcox Solar Observatory, and then they studied the relation of these IMFs with other solar activity indicators.

The EMD is usually applied to decompose a signal $x(t)$ into a series of mono-component contributions designated as IMF and a secular trend (or a constant), namely:

\begin{equation}
x(t)=\sum_{i=1}^{n} s_{i}(t)+r_{n}
\end{equation}

where $r_{n}$ is the residue (either a trend or a constant) after the $n$ IMFs are extracted. Generally speaking, the first IMF $s_1$ contains the shortest periodic scale of the original signal. For an IMF, it should satisfy two conditions: (1) in the whole data set, the number of extrema and the number of zero crossings must either equal or differ at most by one; and (2) at any point, the mean value of the upper envelope and lower envelope is zero \citep{2013Ap&SS.343...27D,2015RAA....15..879Q,2017MNRAS.472.2913G}.

The major drawbacks of the EMD is the frequent appearance of mode mixing, which is defined as a single IMF either consisting of signals of widely disparate scales or a signal of a similar scale residing in different IMFs. To deal with the mode mixing problem, a noise-assisted data analysis method called the ensemble EMD (EEMD) is proposed by \cite{2009AADA...1...1}. The EEMD, which defines the true IMFs as the mean of an ensemble of trials, each consisting of the signal plus a white noise of finite amplitude. This technique is based on the insight gleaned from statistical studies of the inherent properties of white noise, which indicated that it is inspired by the noise-added analysis initiated by \cite{2004RSPSA.460.1597W}. In our analysis, this powerful approach is applied to decompose the solar time series.

\subsubsection{Cross-recurrence Plot}

In the last three decades, the technique of recurrence plot (RP) has been considered as a method to describe the complex dynamics of the nonlinear and non-stationary systems \citep{1987EL......4..973E}. A RP is a representation of recurrent states of a dynamical system in its $m$-dimensional phase space \citep{2007PhR...438..237M}. From a mathematic point of view, it is a pairwise test of all phase space vectors $\vec{x}_{i}\left(i=1, \ldots, N, \vec{x} \in \mathcal{R}^{m}\right)$ among each other, whether or not they are close:

\begin{equation}
R_{i, j}=\Theta\left(\varepsilon-d\left(\vec{x}_{i}, \vec{x}_{j}\right)\right)
\end{equation}

where $\Theta(\cdot)$ is the Heaviside function, $\mathcal{E}$ is a threshold for proximity, and the closeness $d\left(\vec{x}_{i}, \vec{x}_{j}\right)$ could be measured in different ways by using spatial distance or local rank order \citep{2009PhLA..373.4246M}. 

To determine the dependencies between two different systems, the cross-recurrence plot (CRP) is introduced and can be considered as a bivariate extension of the RP \citep{1998PhLA..246..122Z}. An important advantage of the CRP is that it could be used to reveal the local differences of the dynamical evolution of close trajectory segments, represented by bowed lines \citep{2002PhLA..302..299M}. A time dilatation or time compression of trajectories leads to a distortion of the diagonal lines, showing the inner relationship between the slope of RP lines and local temporal transformations \citep{2007PhR...438..237M}. When the two systems are different, the main black diagonal will become somewhat disrupted and is named as line of synchronisation (LOS). Therefore, the LOS allows us to find the phase shift between two time series \citep{2017JSWSC...7A..34D}.

   \begin{figure}
   \centering
   \includegraphics[width=14.0cm, angle=0]{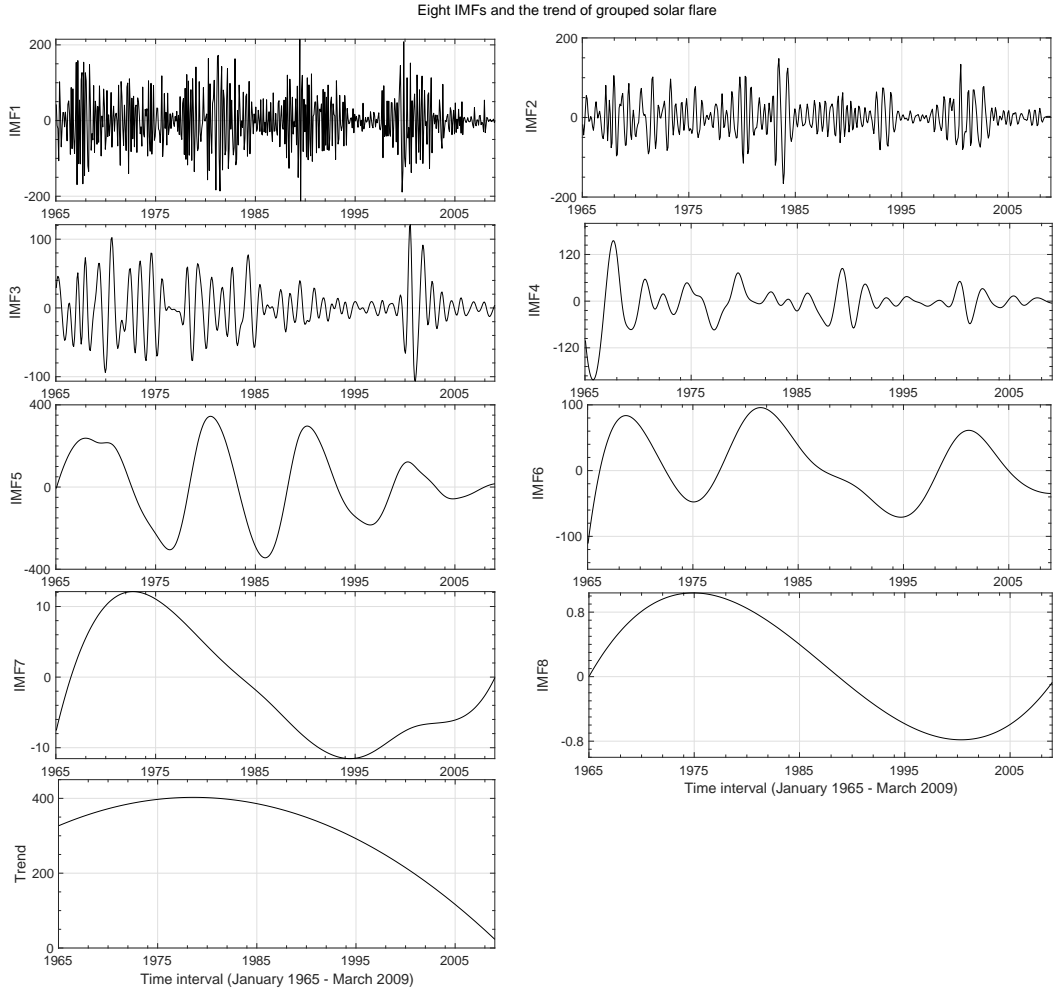}
  % \begin{minipage}[]{85mm}
   \caption{The IMFs 1-8 and the trend of GSF generated by the EEMD method (with the ensemble size of 100). }
%\end{minipage}
   \label{Fig2}
   \end{figure}

\section{Statistical analysis results}

\subsection{Periodic variation of GSF and SN}

   \begin{figure}
   \centering
   \includegraphics[width=14.0cm, angle=0]{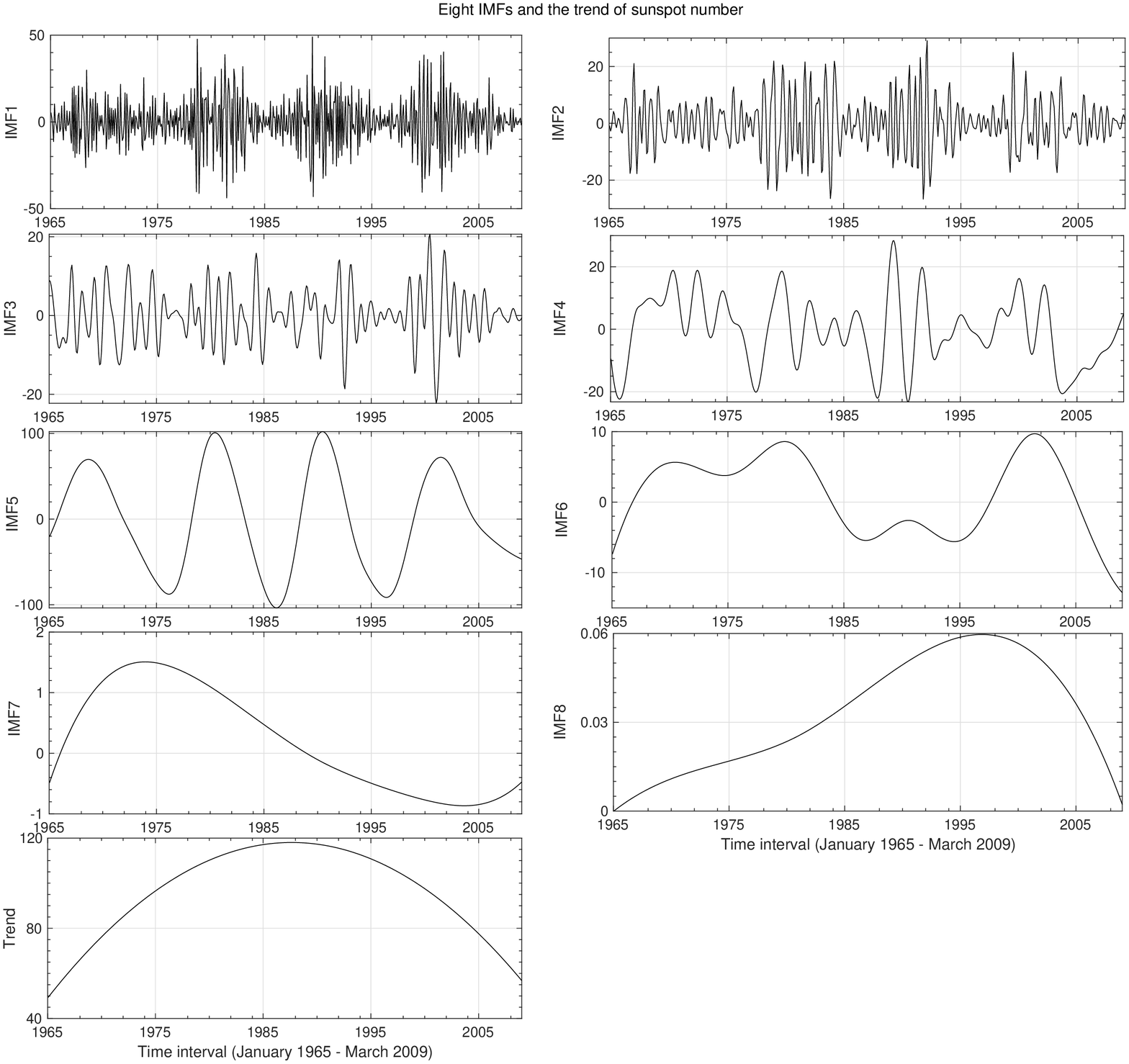}
  % \begin{minipage}[]{85mm}
  \caption{The IMFs 1-8 and the trend of SN generated by the EEMD method (with the ensemble size of 100). }
%\end{minipage}
   \label{Fig3}
   \end{figure}

For the EEMD method, the number of ensemble and the noise amplitude are the two parameters that are needed to be prescribed. As pointed out by \cite{2009AADA...1...1}, an ensemble number of a few hundred could bring out a good result, and the remaining noise would lead to only less than a fraction of 1\% of error if the added noise has an amplitude that is a fraction of the standard deviation of the original time series. In our analysis, the number of ensemble is 100, and the added noise amplitude is 0.2 times of the standard deviation of the original data.

Because the EEMD method is a powerful algorithm to isolate signals with specific timescales in a given time series produced by different underlying physics \citep{2012MNRAS.423.3584L,2017SoPh..292..124G}, it is thus applied to extracted the IMFs of GSF and SN data sets. The decomposed results of GSF and SN, respectively, are shown in Figures 2 and 3. From these two figures, one can easily see that both data sets are decomposed into eight IMFs and a secular trend.

Obviously the extracted IMFs of GSF and SN have time-dependent amplitudes and differ from pure sinusoidal functions. Actually, they are the intrinsic fluctuations decomposed directly from the time series using the shifting process and are pre-estimated functions. However, the IMFs capture the oscillations of the time series even though the data set is not-stationary and nonlinear. For example, the 11-year Schwabe cycle is clearly shown as the fifth extracted IMF in Figures 2 and 3.

Subsequently,  to calculate the average periodicity and the uncertainty of each IMF for GSF and SN, the statistical significance test method proposed by \cite{2004RSPSA.460.1597W} is applied. In this work, we selected two confidence-limit levels: 95\% and 99\%. Namely, the IMFs whose energy level lies above the spread lines, for the white noise, are statistically significant at 95 and 99 percent confidence levels. The uncertainty in the average periodicity is calculated from the standard errors ($\sigma$/$n^{1/2}$, where $\sigma$ is the standard deviation and $n$ is the number of data points). 

Figure 4 displays that for both GSF and SN, IMF1, IMF4, IMF5, and IMF6 are statistically significant at the 99\% confidence level, and the other IMFs are below the 95\% confidence level. Here, the periodicities those are below the 95\% confidence level are not discussed, we only analyzed IMF1, IMF4, IMF5, and IMF6. The average periodicities and the uncertainties of each IMF for GSF and SN are collected in Table 1. As shown in Figure 4 and Table 1, the periodicity (0.2405 year, about 87.7 days) of IMF1 for both data sets is inferred to be three multiple harmonics of 29 days, which is approximately the periodicity of the differential rotation periodicity of the Sun \citep{2014AJ....148...12X}. The average periodicity of IMF4 for each data set is related to the typical timescale between 1 and 4 years, one is 2.5731$\pm$0.1775 years, and the other one is 3.1350$\pm$0.5624 years, they could be considered as the 
solar quasi-biennial oscillation (QBO). The physical origin of solar QBO may be related to the dynamic processes in the solar tachocline, although it is not yet fully understood \citep{2014SSRv..186..359B}. The IMF5 values of GSF and SN should be the most prominent periodicities: the 11-year solar activity cycle, the so-called eleven-year Schwabe cycle (ESC). The periods of IMF6 are 14.865 years for GSF and 17.221 years for SN, respectively, and they are about 1.5 times (1.5$\times$11 years=16.5 years) as long as the 11-year Schwabe cycle. Therefore, the significant periodicities of GSF and SN, those are above the 99\% confidence level, are connected with the differential rotation periodicity, the quasi-biennial oscillation, and the 11-year Schwabe cycle.

     \begin{figure}
   \centering
   \includegraphics[width=14.0cm, angle=0]{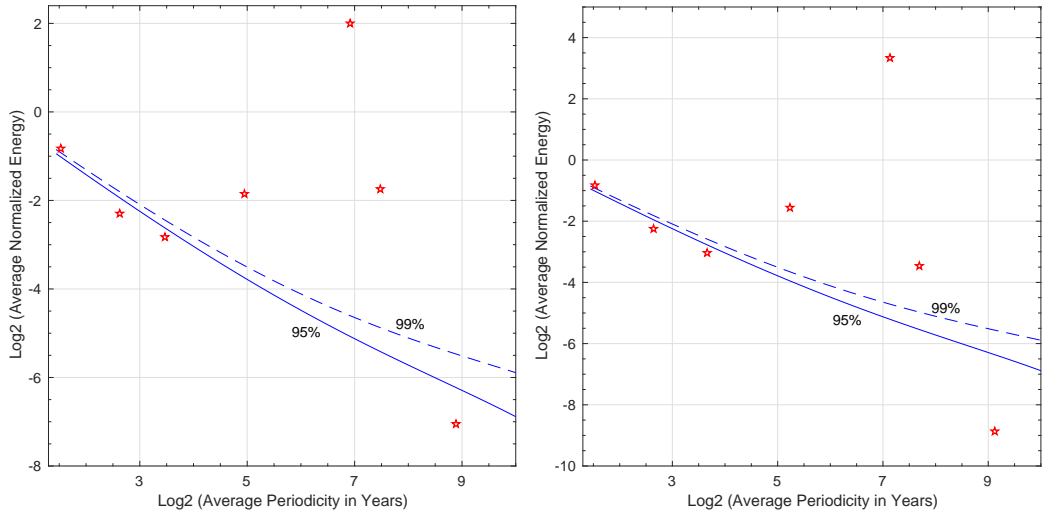}
  % \begin{minipage}[]{85mm}
   \caption{Statistical significance test of seven IMFs of GSF (left panel) and SN (right panel) data sets, respectively. The dashed and the solid lines shown in each panel are the 99\% and 95\% confidence levels.}
%\end{minipage}
   \label{Fig4}
   \end{figure}

\subsection{Phase relationship between GSF and SN}

   \begin{table}
\bc
\begin{minipage}[]{100mm}
\caption[]{The average periodicities (in years) and the uncertainties of IMFs, which are extracted from GSF and SN, respectively. \label{mbh}}\end{minipage}
\setlength{\tabcolsep}{7.5pt}
\small
 \begin{tabular}{ccccccccccccc}
  \hline\noalign{\smallskip}
  &  grouped solar flare & sunspot number  \\
  \hline\noalign{\smallskip}
1 & 0.2405$\pm$0.0032 &  0.2405$\pm$0.0035 \\
2 & 0.5145$\pm$0.0148 &  0.5206$\pm$0.0169 \\
3 & 0.9219$\pm$0.0300 &  1.0536$\pm$0.0219 \\
4 & 2.5731$\pm$0.1775 &  3.1350$\pm$0.5624 \\
5 & 10.069$\pm$0.4102 &  11.698$\pm$0.4169 \\
6 & 14.865$\pm$0.6325 &  17.221$\pm$1.8170 \\
7 & 39.477$\pm$5.3529&  46.484$\pm$4.7843 \\
  \noalign{\smallskip}\hline
\end{tabular}
\ec
%% place \tablecomments and \tablerefs below \end{center| and \end{center}:
%% you may leave the table-width parameter to editors or set to your actual size
%\tablecomments{0.86\textwidth}{
%Black hole masses estimated with Eq. (7) in \cite{Vestergaard06}.}
\end{table}

    \begin{figure}
   \centering
   \includegraphics[width=14.0cm, angle=0]{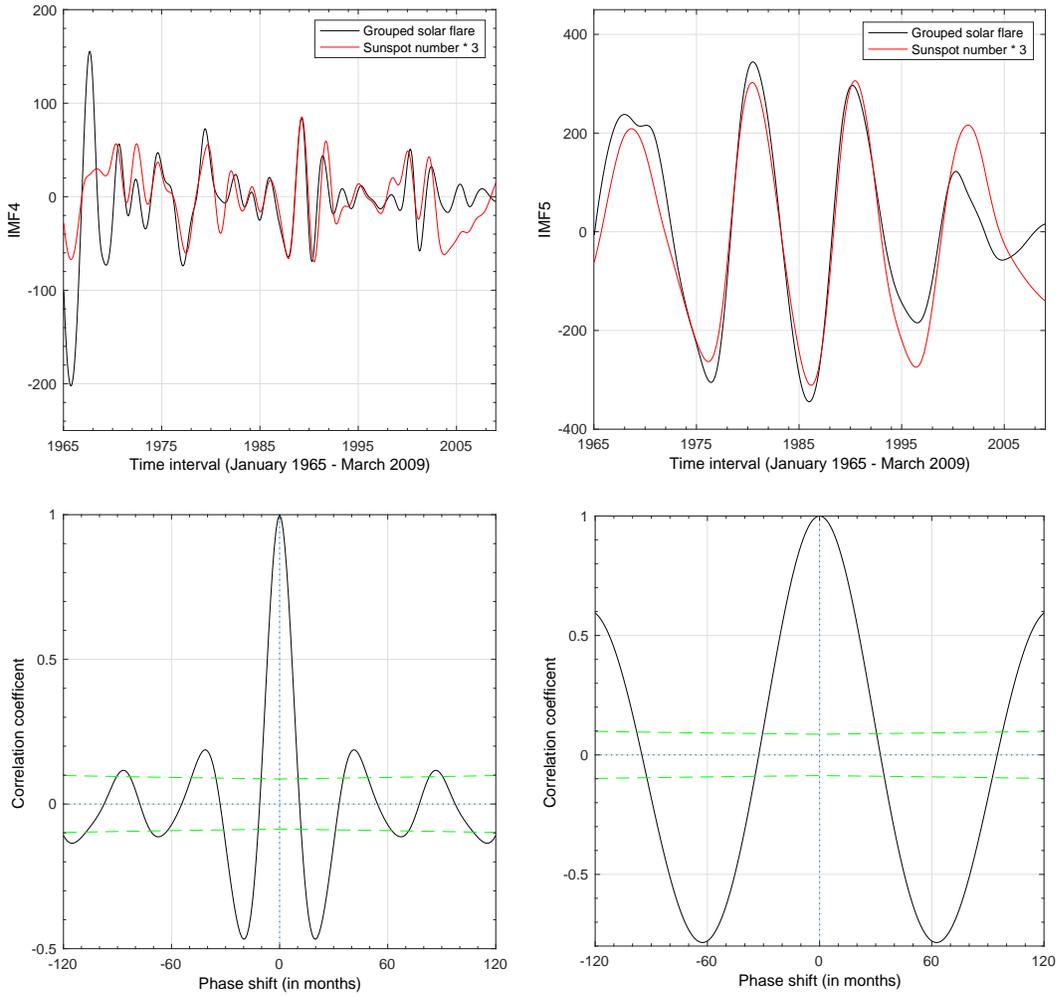}
  % \begin{minipage}[]{85mm}
   \caption{Upper-left panel: temporal variation of QBO signal of GSF (black line) and SN (red line). Upper-right panel: temporal variation of ESC signal of GSF (black line) and SN (red line). Lower-left panel: the cross-correlation analysis results of the QBO signal between GSF and SN, and the dashed green lines are the 95\% confidence levels. Lower-right panel: the cross-correlation analysis results of the ESC signal between GSF and SN, and the dashed green lines are the 95\% confidence levels}
%\end{minipage}
   \label{Fig5}
   \end{figure}
   
        \begin{figure}
   \centering
   \includegraphics[width=14.0cm, angle=0]{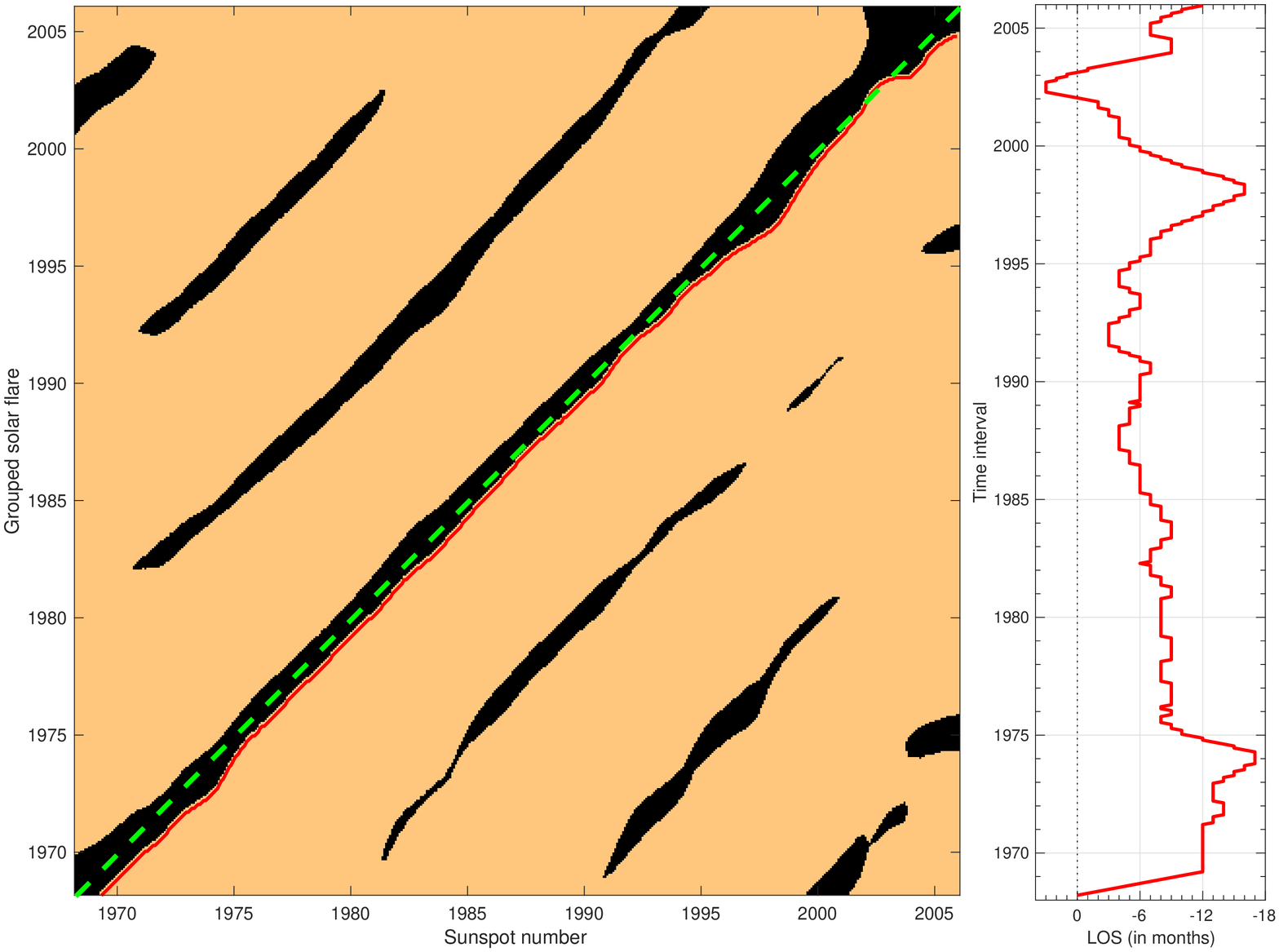}
  % \begin{minipage}[]{85mm}
   \caption{Left panel: the CRP pattern between GSF and SN for the ESC signal, the dashed green line and the solid red line are the main diagonal and the LOS, respectively. Right panel: the LOS, which is used to reveal the local phase difference, extracted from the CRP pattern.}
%\end{minipage}
   \label{Fig6}
   \end{figure}

Although the previous studies showed that solar flare activity lags behind the sunspot number (or area) with several to tens of months, these studies did not consider the different periodic scales that are responsible for the phase difference. Here, two periodic scales are separately studied the phase relationship between GSF and SN, one is the QBO, and the other one is the ESC. To better understand their phase relationship, the cross-correlation analysis method is firstly used, and then the CRP technique is applied.

The upper panels of Figure 5 show the temporal variations of QBO (upper-left) and ESC (upper-right) signals of GSF and SN, respectively. From this figure one can see that their variabilities are not completely in phase, implying that they should be asynchronous in the given periodic scale. The lower panels of Figure 5 display the results of the cross-correlation analysis of QBO (lower-left) and ESC (lower-right) signals of GSF and SN with the phase shifts from -120 to 120 months. The abscissa implies the phase lag of GSF with respect to SN along the calendar-time axis, with the negative values representing the backward shifts. In our analysis, the phase lags are only considered between -120 and 120 months, so all of the local peaks are above the 95\% confidence levels (showed by the dashed green lines).

For the QBO signal, the correlation coefficient is 1 when there is no phase shift, implying that they are highly positive correlation. When the phase shifts are -87, -41, 41, and 87 months, the values of the correlation coefficient arrive at local maxima. The average interval between the two neighboring local maxima is 43.5$\pm$2.9 months (3.63$\pm$0.24 years). When the phase lags are -115, -67, -20, 20, 67, and 115 months, the values of the correlation coefficient reach local minima. The average interval between the two neighboring local minima is 46$\pm$3.4 months (3.83$\pm$0.28 years). For the ESC signal, the situation is very simple. Two signals are highly positive correlation, because the correlation coefficient is 1 when there is no phase shift. When the phase lags are -63 and 63 months, the values of the correlation coefficient arrive at local minima, and the interval between these two neighboring local minima is 126 months (10.5 years).

Based on the cross-correlation analysis, both of the QBO and the ESC signal between GSF and SN are highly positive correlation. However, this method is usually used to measure the long-term variation between different solar activity indicators, causing an averaging of the property being measured, as pointed out by \citep{2013RAA....13..104D}. To compare the timescales of the two data series on a point-by-point basis, another useful tool, the so-called CRP approach can be used. In our analysis, we also apply this powerful technique to reveal their phase relationship for the ESC signal.

The most important advantage of the CRP approach is that the local dependency described by bowed lines of two data sets can be determined and estimated. To construct the CRP pattern between GSF and SN for the ESC signal, the embedding dimension of three, which was calculated by many authors, is used. Fox example, \cite{2006A&A...449..379L} studied the dynamical behavior of sunspot activity by the nonlinear theory, and found that an embedding dimension equal to three is sufficient to reveal its dynamical structure. To investigate the nonlinear dynamical behavior of the polar faculae and sunspot activity for the time interval from 1951 August to 1998 December, \cite{2016AJ....151....2D} used several nonlinear time series analysis approaches, and found that both the high- and the low-latitude solar activity are governed by a three-dimensional chaotic attractor. Another important parameter for constructing the CRP pattern is the time delay, which was estimated to be varying from 29 to 31 when different indicators used \citep{2016AJ....151....2D}. Here, this value of 30 is used in our analysis. The left panel of Figure 6 displays the CRP pattern between GSF and SN for the ESC signal, with the dashed green line and the solid red line showing the main diagonal and the LOS, respectively. Here the LOS could be applied to recognize which time series leads or lags in phase. From this panel, one can easily see that the LOS lies close to the main diagonal, but it still has some deflection from the main diagonal at many data points. We extracted the LOS from the CRP pattern, and showed it in the right panel of Figure 6. If the LOS is greater (smaller) than zero, the GSF will lead (lag) the SN in phase. We found that most of the LOS values are smaller than zero, and the average of all LOS values is -7.8 months. That is, the ESC signal of GSF lags in phase with 7.8 months during the considered time interval. Our analysis result is agreement and further enhances the results obtained by previous studies. 

For the QBO signal, we also studied their phase relationship by the CRP technique, but no such regularity is found. Therefore, we can arrive at a conclusion that the systematic phase delays between GSF and SN originate from the inter-solar-cycle signal (the so-called ESC signal).

\section{Discussion}
\label{sect:discussion}

With the data sets of GSF and SN for the time interval from January 1965 to March 2009, their periodic variation and phase relationship were studied by two nonlinear time-frequency analysis techniques. First, the time series of GSF and SN were decomposed into eight IMFs and a trend through EEMD technique. Second, the significant periodicities of the first seven IMFs were studied by the significance test method. And last, the CRP approach is applied to reveal their phase relationship at different timescales. 

Based on the EEMD analysis, the extracted IMFs of both GSF and SN have time-dependent amplitudes and differ from pure sinusoidal functions. Furthermore, the significant periodicities of both GSF and SN, those are above the 99\% confidence level, are connected with the differential rotation periodicity, the QBO signal, and the ESC signal. However, the specific values of these periodicities are not absolutely identical. From the CRP analysis, it is found that the ESC signal of GSF lags behind (the LOS values smaller than zero) that of SN with an average value of 7.8 months during the considered time interval, but for the QBO signal, no such systematic regularity was found. That leads us to conclude that the systematic phase delays between GSF and SN originate from the ESC signal.

There is no doubt that both GSF and SN have similar quasi-periodic features, because they have intrinsic link with the solar magnetic field. However, they are related in various ways to different aspects of magnetic processes taking place on the Sun (one is in the photosphere, and the other one is in the corona), so their long-term variations differ in fine details. \cite{2000ApJ...540..583S} studied the relationship between  sunspots and large flares, and found that there is a general trend for large regions to produce large flares, but it is less significant than the dependence on magnetic class. \cite{2017MNRAS.465...68E} analyzed different types (C, M, and X classes) of X-ray solar flares occurring in sunspot groups (a total of 4262 active regions) for the time period 1996-2014, and found that large and complex sunspot groups have the flare-production potential about eight times higher than the small and simple active regions. To understand the periodic variations and distributions of solar flares with the sunspot group numbers between 1996 July and 2016 December, \cite{2019ApJ...874...20O} studied the periodicities and distributions of the solar soft X-ray flares with B, C, M, and X-class. They found that the difference in periodic variations of the flare classes could be attributed to the magnetic flux system of sunspot groups producing them. \cite{2011ApJ...731...30K} analyzed the solar activity cycle by focusing on time variations of the number of sunspot groups as a function of their modified Zurich class, and found that large sunspot groups appear to reach their maximum in the middle of the solar cycle (phases 0.45-0.5), while the international sunspot numbers and the small sunspot groups generally peak much earlier (phases 0.29-0.35). we infer that it is also a possible reason for explaining the phase different (tens of months) between flare activity and sunspot activity, which has been confirmed by \cite{2013NewA...23....1D} who studied the relative phase analyses of 10.7 cm solar radio flux with sunspot numbers.

Concluding, in this paper we showed that GSF is delayed with regard to SN with an average of 7.8 months for the period 1965-2009. The obtained phase lag is smaller than the characteristic value (in the range between 10 and 15 months) derived by \cite{2003SoPh..215..111T}.  To understand the phase relationship between GSF and SN at different timescales, the energy balance in the flaring solar corona should be considered. From a model of the dynamic energy balance in the corona over the solar cycle, \cite{2001ApJ...557..332W} found that the expected global coronal response time (the time for flares to remove available coronal energy) is 8.8 months. Solar flare is a sudden release of magnetic energy that was stored in the corona, and the flare rate is believed to increase with the stored energy. The available magnetic energy are expected to vary cyclically with the rate of energy supply to the corona, but with a phase delay with respect to the variation in energy supply. Therefore, our analysis result is consistent with and further enhances the numerical model proposed by  \cite{2001ApJ...557..332W}, and may provide evidence about the storage of magnetic energy in the corona.

\normalem
\begin{acknowledgements}

This research uses sunspot data obtained from the WDC-SILSO, Royal Observatory of Belgium, Brussels (http://www.sidc.be/silso/). This work is supported by the National Key Research
and Development Program of China (2018YFA0404603), the Joint Research Fund in Astronomy (Nos. U1631129, U1831204,U1931141) under cooperative agreement between the National Natural Science Foundation of China (NSFC) and Chinese Academy of Sciences (CAS), the fund of the National Natural Science Foundation of China (11903009), the Yunnan Key Research and Development Program (2018IA054), the open research program of CAS Key Laboratory of Solar Activity (Nos. KLSA201807), and the major scientific research project of Guangdong regular institutions of higher learning (2017KZDXM062).

\end{acknowledgements}

\bibliographystyle{raa}
\bibliography{raawang.bib}

\end{document}